\documentclass[runningheads]{llncs}
\usepackage[T1]{fontenc}
\usepackage{graphicx}
\usepackage{comment}
\usepackage{soul}
\usepackage{hyperref}
\usepackage{xcolor}
\usepackage{algorithm}
\usepackage{algpseudocode}
\usepackage{amsmath}
\usepackage{fontawesome5}

\usepackage{tikz}
 
\newcommand*\halfcirc[1][1ex]{%
  \begin{tikzpicture}
  \draw[fill] (0,0)-- (90:#1) arc (90:270:#1) -- cycle;
  \draw (0,0) circle (#1);
  \end{tikzpicture}}

\newcommand*\fullcirc[1][1ex]{\tikz\fill (0,0) circle (#1);}

\begin{document}
\title{ICSSPulse: A Modular LLM-Assisted Platform for Industrial Control System Penetration Testing}
%
%
\author{Michail Takaronis\inst{1}\orcidID{0009-0007-2593-8811}\faIcon{envelope} \and
Athanasia Kollarou\inst{1}\orcidID{0009-0008-8239-4432} \and
Vyron Kampourakis\inst{1}\orcidID{0000-0003-4492-5104} \and 
Vasileios Gkioulos\inst{1}\orcidID{0000-0001-7304-3835} \and
Sokratis Katsikas\inst{1}\orcidID{0000-0003-2966-9683}}
\authorrunning{M. Takaronis et al.}
%
\institute{Norwegian University of Science
and Technology, 2802 Gjøvik, Norway \email{\{michail.takaronis, athanasia.kollarou, vyron.kampourakis, vasileios.gkioulos, sokratis.katsikas\}@ntnu.no}}
\maketitle              
\begin{abstract}
It is well established that industrial control systems comprise the operational backbone of modern critical infrastructures, yet their increasing connectivity exposes them to cyber threats that are difficult to study and remedy safely under real-time operational conditions. In this paper, we present \textit{ICSSPulse}, an open-source, modular, and extensible penetration testing platform designed for the security assessment of ICS communication protocols. To the best of our knowledge, \textit{ICSSPulse} is the first web-based platform that unifies network scanning, protocol-aware Modbus and OPC~UA interaction, and Large Language Model (LLM)-assisted reporting within a single, lightweight ecosystem. Our platform provides a user-friendly graphical interface that orchestrates enumeration, exploitation, and reporting activities over simulated industrial services, enabling safe and reproducible experimentation. It supports protocol-level discovery, asset enumeration, and controlled read/write interactions, while preserving protocol fidelity and operational transparency. Experimental evaluation using synthetic Modbus test servers, a \textit{Factory I/O} water treatment scenario, and a custom OPC~UA production-line model demonstrated \textit{ICSSPulse's} potential to discover active industrial services, enumerate process-relevant assets, and manipulate process variables. A key contribution of this work lies in the integration of an LLM-assisted reporting module that automatically translates technical findings into structured executive and technical reports, with mitigation guidance informed by the ICS MITRE ATT\&CK ICS matrix.

\end{abstract}
\section{Introduction}
\label{S:intro}

Industrial Control Systems (ICS) comprise the operational backbone of modern Critical Infrastructures (CI), including energy grids, water treatment plants, transportation networks, and manufacturing plants. Based on the PURDUE model~\cite{PURDUE}, these systems integrate Operational Technology (OT) components at their lower layers, spanning supervisory control like Supervisory Control and Data Acquisition (SCADA) systems, field control including Programmable Logic Controllers (PLCs) and Remote Terminal Units (RTUs), and field devices such as sensors and actuators. At the same time, these layers are vertically connected with Information Technology (IT) networks, enabling process automation and real-time control. While this convergence enhances operational efficiency, it also exposes traditionally isolated industrial networks to cybersecurity threats previously unknown to the once barricaded industrial operations~\cite{makr2021,KAMPOURAKIS2025104673}.

The consequences of compromising ICS environments are no longer hypothetical. For example, the Stuxnet worm (2010)~\cite{falliere2011w32}, one of the earliest known cyber-physical attacks, demonstrated the feasibility of stealthily modifying PLC logic to cause physical damage, in this case, degrading centrifuge operations at Iran’s Natanz facility. Similarly, incidents such as the TRITON malware (2017)~\cite{di2018triton}, which targeted safety instrumented systems in a Saudi petrochemical plant, and the Industroyer (2016)~\cite{Industroyer}, which disrupted power distribution in Ukraine by directly manipulating industrial communication protocols, underscored the catastrophic potential of ICS-targeted cyberattacks. These events blatantly prove that cybersecurity threats in ICS environments can have direct and far-reaching impacts on economic stability, public infrastructure, and even human safety. Despite their gravity, educating and training the future cybersecurity workforce in ICS security remains an underdeveloped and underfunded domain. However, unlike conventional IT systems, availability and safety are a sine qua non in industrial systems, making aggressive security testing or direct intrusion simulations difficult to perform on live infrastructure. Consequently, there is an increasing need for controlled environments and specialized tools, supporting secure and repeatable testing of ICS protocols and diverse configurations.

In this direction, Penetration Testing (PT) remains a highly relevant hands-on practice. However, in industrial contexts, it differs significantly from that in traditional IT networks. Namely, common IT-focused frameworks, such as Metasploit or Nmap, are not fully equipped to handle the semantics, timing constraints, or deterministic behavior of ICS communication standards such as Modbus, DNP3, and OPC~UA. As a result, testing such protocols requires tools that can safely emulate and interface with industrial devices, capture real-time data exchanges, and assess the security posture of ICS communication channels without risking operational continuity. In this context, aside from large-scale proprietary testbeds~\cite{swat2016,Magnus2019,wadi2017}, existing contributions in the field are scarce and only marginally relevant to our study. However, for a broader perspective, the reader is referred to~\cite{green2017pains,maynard2018open,Xie2018VTET,Kraust2025PT}.

\noindent \textbf{\textit{Contribution:}} Towards filling these gaps, \textit{ICSSPulse}~\cite{ICSSPulse}, an open-source ICS protocol platform, provides a modular and extensible environment designed to facilitate interactive testing, simulation, and visualization of industrial communication protocols. By combining a web-based Graphical User Interface (GUI) with backend modules for Modbus and OPC~UA, along with a novel Large Language Model (LLM)-assisted reporting module, \textit{ICSSPulse} enables safe, reproducible, and protocol-aware PT activities. The platform is particularly suited for research laboratories, cyber ranges integration, and training scenarios, where it can be used to study protocol vulnerabilities, validate security mechanisms, and train operators in recognizing abnormal communication patterns. Through its modular design and integration of simulated data sources, \textit{ICSSPulse} aims to bridge the gap between theoretical ICS cybersecurity analysis and practical, hands-on experimentation.

The remainder of this paper is organized as follows. Section~\ref{S:Threat:Model} presents \textit{ICSSPulse's} scope. Section~\ref{sec:lifecycle} describes the integration of \textit{ICSSPulse} within the PT lifecycle. Section~\ref{S:arch} details \textit{ICSSPulse's} architecture. Section~\ref{S:impl} discusses the implementation of the platform’s core components. Section~\ref{S:eval} presents the experimental evaluation. Last, Section~\ref{S:concl} concludes the paper and outlines directions for future work.

\section{Scope}
\label{S:Threat:Model}

It is common that ICS exposes legacy or weakly protected communications, which, if improperly configured, may leak process values or control information to unauthorized parties. In this context, such exposures directly threaten the confidentiality and integrity of operational parameters and, depending on the adversary's intent and capabilities, may affect availability as well. In the scope of this work, \textit{ICSSPulse} focuses specifically on protocol-level security threats, which can be leveraged by an attacker who targets unauthenticated or weakly authenticated communication paths in network communication implementations such as Modbus or OPC~UA. Note that \textit{ICSSPulse} models communication behavior rather than physical actuation, meaning that threats pertinent to device firmware or hardware-level components currently fall outside its scope. In this mindset, we assume that \textit{ICSSPulse} can be utilized by adversaries that have logical network access to the lower layers of an ICS environment, attained either by compromising an operator workstation, pivoting through a flat network segment, or leveraging improperly segmented infrastructure. Of course, the mechanisms by which this network access is achieved, e.g., spear-phishing, lateral movement, or exploitation of unpatched services, are relevant in real deployments; however, they are currently outside the coverage of \textit{ICSSPulse}. Instead, we focus on the adversary's behavior once they can communicate with industrial endpoints exposed over the industrial network. From this standpoint, the evildoer can perform active probing, enumerate accessible endpoints, and issue protocol-level requests to field devices.

Consequently, once the attackers can interact with relevant services, they can identify exposed registers, memory blocks, or nodes that reveal operational information. For example, in Modbus communications, this may include coils or holding registers, while in OPC~UA, accessible node paths or namespaces. Subsequently, the threat actors can construct requests to read or modify values, test the behavior of endpoints under malformed inputs, or generally assess the field devices' cybersecurity robustness. With \textit{ICSSPulse}, these operations occur on simulated protocol handlers, enabling safe reproduction of adversarial behaviors for analysis, training, and education. In its current version, \textit{ICSSPulse} does not model physical disruptions, controller reprogramming, multi-stage ICS kill chains, or privileged engineering workstation attacks. Similarly, Denial of Service (DoS) targeting device availability, packet flooding, or resource exhaustion attacks are outside the platform's use. Recall that presently \textit{ICSSPulse} provides a controlled environment for studying adversaries operating at the protocol interaction level. Summarizing the key characteristics, goals, and capabilities of the adversary, the scope of \textit{ICSSPulse} can be defined as follows.

\begin{itemize}
    \item \textbf{Goal}: The attacker aims to obtain or manipulate process-related data by interacting with exposed protocol endpoints. This may include reading operational values, issuing unauthorized write commands, or probing device behavior for unveiling weaknesses.

    \item \textbf{Knowledge}: The adversary possesses general knowledge of industrial network protocols, in \textit{ICSSPulse's} case, Modbus and OPC~UA, understanding common addressing schemes or message structures. However, they lack detailed system-specific knowledge, such as device configuration files, engineering project files, or process logic.

    \item \textbf{Technical skills}: The attacker is skilled enough to generate and interpret protocol messages, conduct enumeration, craft malformed requests, and analyze device responses. Skills include basic network probing, familiarity with industrial protocol semantics, and the ability to identify unprotected or weakly protected endpoints.

    \item \textbf{Capabilities}: The adversary can send and receive network traffic to devices using industrial protocols. They can enumerate accessible memory/register spaces, test device behavior under different request types, and exploit unauthenticated protocol functions. However, they cannot compromise firmware, alter physical I/O, or directly attack the hardware. Their capabilities are limited to protocol-level interactions within the network segment to which they have obtained access.
\end{itemize}

\section{Lifecycle Integration}
\label{sec:lifecycle}

Penetration testing is commonly defined as ``a test methodology in which assessors, typically working under specific constraints, attempt to circumvent or defeat the security features of an information system''~\cite{NIST:PT} following a structured six-stage lifecycle. Typically, the sequence of a typical PT pipeline is: i) planning \& reconnaissance, ii) scanning \& enumeration, iii) vulnerability analysis, iv) exploitation, v) post-exploitation, and vi) reporting \& remediation. Accordingly, \textit{ICSSPulse} provides mechanisms and modules that fully or partially support a full PT campaign in a controlled environment. Table~\ref{tab:lifecycle} summarizes \textit{ICSSPulse's} PT lifecycle coverage. 

\noindent \textbf{Planning \& Reconnaissance}: This stage involves defining engagement scope and preparing testing parameters. In this regard, \textit{ICSSPulse} supports this phase by allowing users to configure target IP addresses, ports, protocol modes, and operation types through its web GUI. Note that validation mechanisms are implemented to ensure consistent configuration relevant to real-world planning activities. 

\noindent \textbf{Scanning \& Enumeration}: Scanning identifies live hosts and open services, while enumeration determines protocol-level characteristics and potential points of interaction. In this context, \textit{ICSSPulse} fully supports this stage through an integrated scanning module for host/port discovery. Additionally, \textit{ICSSPulse} exposes protocol-level handlers, performing operations such as coil/register enumeration, unit identifier discovery, and probing of common address ranges.

\noindent \textbf{Vulnerability Analysis}: In this stage, testers analyze the collected data to identify potential weaknesses that could lead to exploitation. \textit{ICSSPulse} facilitates this by presenting structured, transparent output for all scanning and enumeration operations executed in the previous step. This way, users can interpret exposed services or insecure behaviors (e.g., unauthenticated Modbus access). Note that analysis is user-driven, as automated vulnerability scoring is not included in the current version of \textit{ICSSPulse}, but is intended as future work.

\noindent \textbf{Exploitation}: Exploitation involves attempting controlled actions based on identified weaknesses. In this respect, \textit{ICSSPulse} supports exploitation through its protocol handler, i.e., Modbus and OPC~UA, which enable read/write operations on coils and registers, and standard object nodes, respectively, testing malformed or unexpected requests.

\noindent \textbf{Post-Exploitation}: This stage evaluates the impact of successful exploitation. \textit{ICSSPulse} supports it by visually displaying device responses and state changes resulting from protocol manipulation, depending on the case. Therefore, users can examine how simulated devices react to altered values or peculiar requests. It should be noted that system-level post-exploitation activities, such as privilege escalation or lateral movement, are beyond the tool's current scope, but they are considered for future extension.

\noindent \textbf{Reporting \& Remediation}: The final stage documents findings and recommends mitigation steps. \textit{ICSSPulse} fully supports this phase through its novel, integrated LLM-powered reporting engine. The system aggregates stored evidence, including scan results, user interactions, and outputs, into a structured technical report, highlighting mitigation guidance aligned with well-known security frameworks.

\begin{table}[!ht]
\centering
\caption{\textit{ICSSPulse's} Penetration Testing Lifecycle Coverage.}
\label{tab:lifecycle}
\resizebox{0.8\textwidth}{!}{%
\begin{tabular}{cp{7.5cm}c}
\hline
\textbf{PT Stage} & \textbf{Description} & \textbf{Support} \\ \hline

\textit{Planning \& Reconnaissance} & Users configure targets, ports, and protocol parameters & \halfcirc \\ \hline

\textit{Scanning \& Enumeration} & Integrated scanning and protocol-aware enumeration (e.g., Modbus unit IDs, register mapping). & \fullcirc \\ \hline

\textit{Vulnerability Analysis} & Structured outputs enable manual identification of insecure services or behaviors; no automated scoring. & \halfcirc \\ \hline

\textit{Exploitation} & Modbus and OPC~UA read/write and malformed request testing fully supported & \fullcirc \\ \hline

\textit{Post-Exploitation} & Impact observation through state-change feedback; broader system-level post-exploitation not included. & \halfcirc \\ \hline

\textit{Reporting \& Remediation} & Integrated LLM-based report generation with evidence aggregation and mitigation recommendations. & \fullcirc \\ \hline
\end{tabular}}
\end{table}

\section{ICSSPulse Architecture}
\label{S:arch}

\begin{figure}[htbp]
\centering
\includegraphics[width=0.9\textwidth]{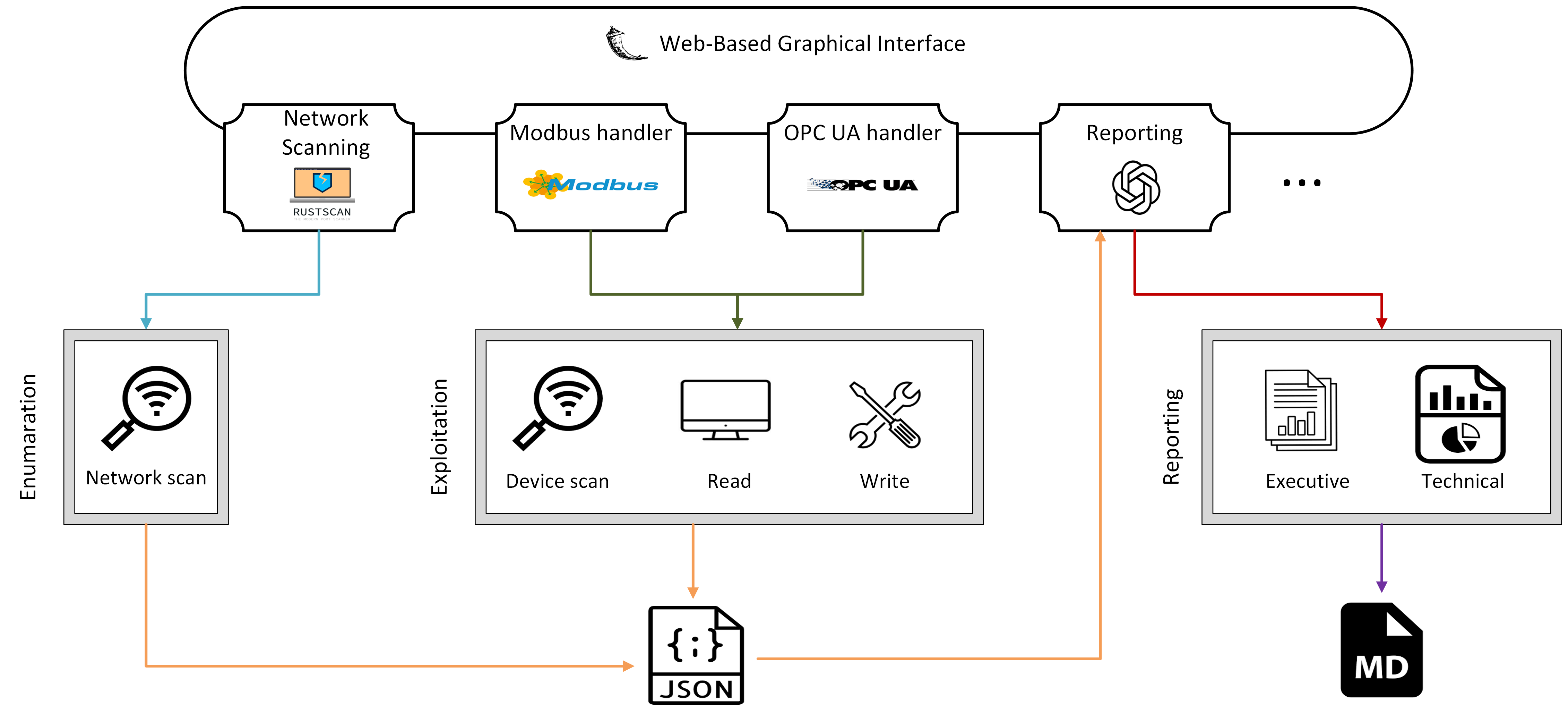}
\caption{ICSSPulse Architecture.}
\label{fig:arch}
\end{figure}

Figure~\ref{fig:arch} illustrates the architecture of \textit{ICSSPulse}, which is designed as a modular, web-based platform for ICS PT. The architecture is centered around a unified web GUI that orchestrates user interactions and coordinates the execution of enumeration, exploitation, and reporting activities, incorporating open-source tools like Rustscan, tailored scripts based on Python libraries like pymodbus and python-opcua, and an LLM agent through a GPT-4o-mini model using OpenAI's API, respectively. This centralized design enables seamless control flow while maintaining a clear separation between user interaction, protocol-specific processing, and result analysis.

As shown in the figure, enumeration and exploitation components operate as independent modules, producing machine-readable outputs in JSON format. These artifacts serve as an intermediate representation, allowing results to be immediately visualized through the GUI or selectively retained for later use. The reporting component takes as input the aggregated artifacts derived from the scanning and enumeration modules to generate executive or technical assessment reports, depending on the users' demands, without direct coupling to the underlying testing modules. Overall, the architecture emphasizes modularity, extensibility, and clear data flow, facilitating reproducible security assessments of ICS protocols.

Architecturally, \textit{ICSSPulse} is designed with extensibility as a core tenet. That is, additional industrial communication protocols (e.g., DNP3, MQTT, or IEC~60870-5-104) can be incorporated by introducing new handler modules and extending the routing logic, without requiring changes to the GUI or existing execution flow. Likewise, \textit{ICSSPulse} can be extended in a full-fledged cyber range ecosystem~\cite{kamp2025}, say, by enhancing the frontend with advanced visualization dashboards, adding real-time monitoring and management, or augmenting the reporting capabilities independently of the protocol-specific components. This modular design makes \textit{ICSSPulse} well-suited for educational and research-oriented cybersecurity applications, enabling safe, transparent, and reproducible cybersecurity assessments in industrial communication environments.

\section{Implementation}
\label{S:impl}

This section describes the implementation details of \textit{ICSSPulse}, focusing on its five core modules: the GUI, the network scanning capabilities, the Modbus and OPC~UA protocol handlers, and the LLM-assisted reporting module.

\subsection{Graphical User Interface}

\textit{ICSSPulse} is implemented using \textit{Flask}, a lightweight Python microframework for web applications. \textit{Flask} was chosen due to its simple and practical architecture, which delivers essential web server functionality, avoiding excessive complexity and unnecessary dependencies. Regarding the application control, we implemented a centralized control node responsible for defining the routing architecture and managing the request–response lifecycle. Specifically, each route corresponds to a dedicated PT module, i) a home interface for navigation and user interaction, ii) a network scanning module for identifying IT/OT devices within the target environment, iii) a Modbus protocol handler that enables Modbus-based operations through a web interface, iv) an OPC UA protocol handler that facilitates OPC~UA interactions via the same interface, and v) a reporting module that leverages the OpenAI API to generate concise, reusable analysis results. All route handlers process HTTPS POST requests containing user-defined parameters. Subsequently, these parameters are passed to the appropriate protocol-specific handler, and the resulting outputs are output to the frontend GUI in a structured and appropriate format for visualization. Note that dynamic content rendering is achieved through the use of \textit{Jinja2} templating, which enables the frontend to remain largely protocol-agnostic while focusing on usability, interaction, and result presentation.

\subsection{Network Scanning}

Concerning network scanning, \textit{ICSSPulse} integrates \textit{RustScan}, a state-of-the-art network scanner implemented in Rust, enabling accurate network discovery. \textit{RustScan} is deployed within a \textit{Docker} container, thereby abstracting dependency management and installation requirements while ensuring consistent behavior regardless of the underlying system. To facilitate this integration, the platform spawns a separate process to execute Docker commands, dynamically launching containers to perform scans and terminating them upon scanning completion. Furthermore, user-provided parameters, including quoted strings and escape sequences, are securely parsed to correctly handle complex command syntax, thus mitigating the risk of command injection vulnerabilities. Subsequently, scan results are captured, with the output prompts including both standard and diagnostic messages, which are forwarded to the web-based QUI. This design enhances the scanning transparency and supports effective debugging. Furthermore, the scanning module optionally comes with a report generation subsystem, allowing users to incorporate scan results into automated reports. Namely, upon user request, the raw scan output, formatted in JSON, is forwarded to the reporting component together with relevant metadata, such as the executed command and timestamp.

\subsection{Modbus handler}
\label{SS:modbus}

The Modbus handler module in \textit{ICSSPulse} is implemented using \textit{pymodbus}, a Python library that provides a full-stack implementation of the Modbus protocol. The library abstracts low-level protocol mechanics while exposing high-level interfaces for interacting with Modbus data structures, enabling read and write operations on elements such as coils and registers. The handler supports all four primary Modbus data types, i.e., (i) coils, which represent single-bit read/write outputs; (ii) discrete inputs, which correspond to single-bit read-only inputs; (iii) holding registers, which store 16-bit read/write values; and (iv) input registers, which contain 16-bit read-only values. The implementation relies on a Modbus TCP client to establish TCP connections with the target devices. Connection parameters, including the IP address, port number (default: 502), unit (slave) identifier, timeout duration, and retry count, are configurable via the GUI. Altogether, the handler supports multiple operational modes tailored to PT and reconnaissance tasks. Particularly, enumeration operations traverse address ranges to retrieve and display all accessible values, thereby facilitating the exploration of device memory maps, while read operations allow targeted retrieval of values from user-specified addresses, with results returned directly to the GUI.

Apart from standard Modbus functionality, the module incorporates device discovery capabilities. Namely, a dedicated unit scanning function probes a range of unit identifiers to discern active Modbus slaves on the target network. For each candidate unit ID, the implementation employs a multi-function probing strategy that tests all four Modbus data types across commonly used address offsets, including 0 (standard starting address), 1 (alternative starting address used by some devices), 100 (common register block offset), 1000 (higher address region), and the Modicon addressing conventions 40001 and 30001 for holding and input registers, respectively. Beyond identifying active devices, the scanner can record which Modbus function codes successfully elicit responses. For example, devices that support only holding registers are identified accordingly, and their supported data types are reflected in the output GUI. Similarly, the register range scanning functionality performs memory enumeration by reading address ranges in configurable chunks (default size: 1000 registers), enabling the identification of accessible memory regions and their current values.

\subsection{OPC~UA handler}
\label{SS:opcua}

On the other hand, the OPC~UA handler leverages \textit{python-opcua}, a library that provides an implementation of the OPC~UA specification, offering programming interfaces for OPC~UA functionalities, such as node discovery and value manipulation. Unlike Modbus, OPC~UA employs a node-based information model, where each node represents an asset, variable, or method within a namespace structure. Thus, the handler supports exploration of the user-given namespace through discovery mechanisms by implementing five core operational modes. First, it initiates an OPC~UA session, offering user authentication and anonymous interaction to retrieve available endpoints from the target server. Then, for each endpoint, the handler lists the endpoint URL, security policy (e.g., Basic256Sha256, None), security mode (e.g., None, Sign, SignAndEncrypt), and supported user identity token types (e.g., Anonymous, Username/Password, Certificate, IssuedToken). Subsequently, the handler performs a recursive traversal of the OPC~UA namespace starting from the standard object nodes, with parameters such as recursion depth and number of nodes configurable through the GUI. For each node, it logs the node identifier (e.g., ns=2; i=10;), name, and namespace index (e.g., 2: Temperature). Afterwards, the handler performs an enumeration of selected variables within the namespace, with the correct filtering based on the namespace index. Then, for each variable node, it captures the node identifier, display name, data type, access levels referring to user permissions (read/write/execute), and the current value of the node. Finally, the handler can read and write on specified nodes by their NodeId, returning the current value for read operations and a message indicating whether write operations were successful or failed through the GUI.

\subsection{LLM-Assisted Report}
\label{SS:report}

Finally, \textit{ICSSPulse} integrates an LLM-assisted reporting module, automatically generating structured assessment reports, as described in~\ref{alg:llm_report}. Note that the modules offer two reporting modes, i.e., executive and technical, depending on the intended audience. Both modes rely on an LLM, namely, GPT-4o-mini via an OpenAI API key, to summarize the collected output from the scanning and Modbus and OPC~UA exploitation steps and generate a cohesive report with mitigation recommendations informed by the ICS MITRE ATT\&CK framework. The reporting module is organized in two distinct stages. First, the platform maintains a centralized inbox where results from network scans and Modbus and OPC~UA interactions are added during the PT campaign. Each stored entry contains the operation parameters, timestamp, and resulting output, allowing relevant contextual information to be preserved without requiring manual management of intermediate data. In the second stage, upon report request, \textit{ICSSPulse} processes the collected data, extracting contextually meaningful data: scan outputs are analyzed to identify hosts, open ports, and exposed protocols, while Modbus interactions are summarized to capture information such as discovered unit identifiers, accessible registers, and indicators of operational success. Ultimately, these elements are aggregated per target in a comprehensive overview.

Subsequently, mitigation recommendations are generated by mapping the resulting report to predefined ICS ATT\&CK matrix mitigation identifiers. Rather than distributing mitigations throughout the report, the reporting module consolidates them into a single adhesive section, ensuring a clear and concise presentation of the recommended security controls. Once this reporting step is concluded, the backend invokes the OpenAI GPT-4o-mini API to construct a formatted dataset containing the extracted findings, target summaries, and selected mitigations, as seen in~\ref{alg:llm_report}. This dataset, together with the instructions corresponding to the selected reporting mode, is provided as input to the LLM to produce the final report. Recall that the reporting module is explicitly constrained to operate on the supplied data, preventing the inclusion of unsupported findings. In executive mode, the generated report provides a high-level overview suitable for decision-makers, whereas the technical mode provides a more detailed and structured report intended for analysts and practitioners. Last, the resulting report is rendered in \textit{Markdown} format through the GUI and can be downloaded.

\begin{algorithm}
\caption{LLM-Assisted Report Generation.}
    \label{alg:llm_report}
    \scriptsize
    \begin{algorithmic}
        \State \textbf{Input:} Audience type $a \in \{\text{executive}, \text{technical}\}$, report title $T$, model identifier $M$
        \State \textbf{Output:} Markdown report $R$
        \State Initialize an empty report inbox $\mathcal{I} \leftarrow [\,]$
        \Statex

        \State \textbf{During testing:} For each tool execution (scan, Modbus, OPC~UA)
        \State \hspace{1em} Collect parameters $p$ and textual output $o$
        \State \hspace{1em} Append item $\langle t, \text{category}, p, o \rangle$ to $\mathcal{I}$ with current timestamp $t$
        \Statex

        \State \textbf{When report is requested:}
        \State Select at most $N$ items $\mathcal{I}' \subseteq \mathcal{I}$ for processing
        \State Derive per-target scan facts (hosts, ports, services) from $\mathcal{I}'$
        \State Derive per-target Modbus summaries (unit IDs, registers, success flags) from $\mathcal{I}'$
        \State Map observed services and Modbus actions to ICS ATT\&CK mitigations
        \State Deduplicate mitigations by identifier to obtain a consolidated list
        \Statex

        \If {$a = \text{executive}$}
            \State Build dataset $D$ with high-level target statistics and consolidated mitigations
            \State Set instruction string $G \leftarrow$ ``Write an executive ICS/OT security report.''
        \Else
            \State Build dataset $D$ with per-target details, Modbus traces, and consolidated mitigations
            \State Set instruction string $G \leftarrow$ ``Write a structured technical ICS/OT report.''
        \EndIf
        \Statex

        \State Serialize $D$ into a JSON-like structure
        \State Construct prompt using $T$, $G$, and the serialized dataset 
        \State Query the LLM with model $M$ to obtain a Markdown report $R$
        \State \textbf{Return:} $R$
    \end{algorithmic}
\end{algorithm}

\section{Evaluation}
\label{S:eval}

Regarding the Modbus handler evaluation, we utilized two distinct scenarios. The first was conducted using a Python-based Modbus TCP server implemented with the \textit{pymodbus} library, as detailed in Section~\ref{SS:modbus}. The server simulated three distinct industrial devices. Specifically, the device with unit ID~1 simulated a PLC exposing 1,000 addresses per Modbus data type, including coils, discrete inputs, holding registers, and input registers populated with random values. Second, the device with unit ID~5 simulated a sensor with monitoring and measurement capabilities, with discrete inputs set to \textit{True} and the remaining inputs set to \textit{False}, while holding registers were populated with values representing analog measurements. Finally, the device with unit ID~10 simulated an actuator, featuring coils divided into two regions, i.e., 500 enabled and 500 disabled, to represent mixed operational states, and holding registers offset by 10,000 to simulate control signal baselines.

The second Modbus evaluation was performed using \textit{Factory I/O}~\cite{factoryIO}, a 3D industrial simulator designed for practicing real-world control tasks. The selected scenario simulated a part of a water treatment plant and was evaluated using both analog and digital signal representations. Particularly, the simulated plant consisted of a water tank with two primary control loops, namely, an inlet fill valve and an outlet discharge valve, managed by sensors that periodically sampled the water level and flow rates. Accordingly, the flow meter and level meter were configured as read-only input registers, returning analog values for flow rate and tank level, respectively. Meanwhile, the fill and discharge valves were implemented as read/write holding registers, controlling inlet and outlet flow and reporting opening percentages ranging from 0 to~10. For digital evaluation, the analog registers were replaced with discrete inputs for the sensors and coils for the actuators, enabling a binary representation of the plant behavior.

\begin{figure}[htbp]
\centering
\includegraphics[width=0.8\textwidth]{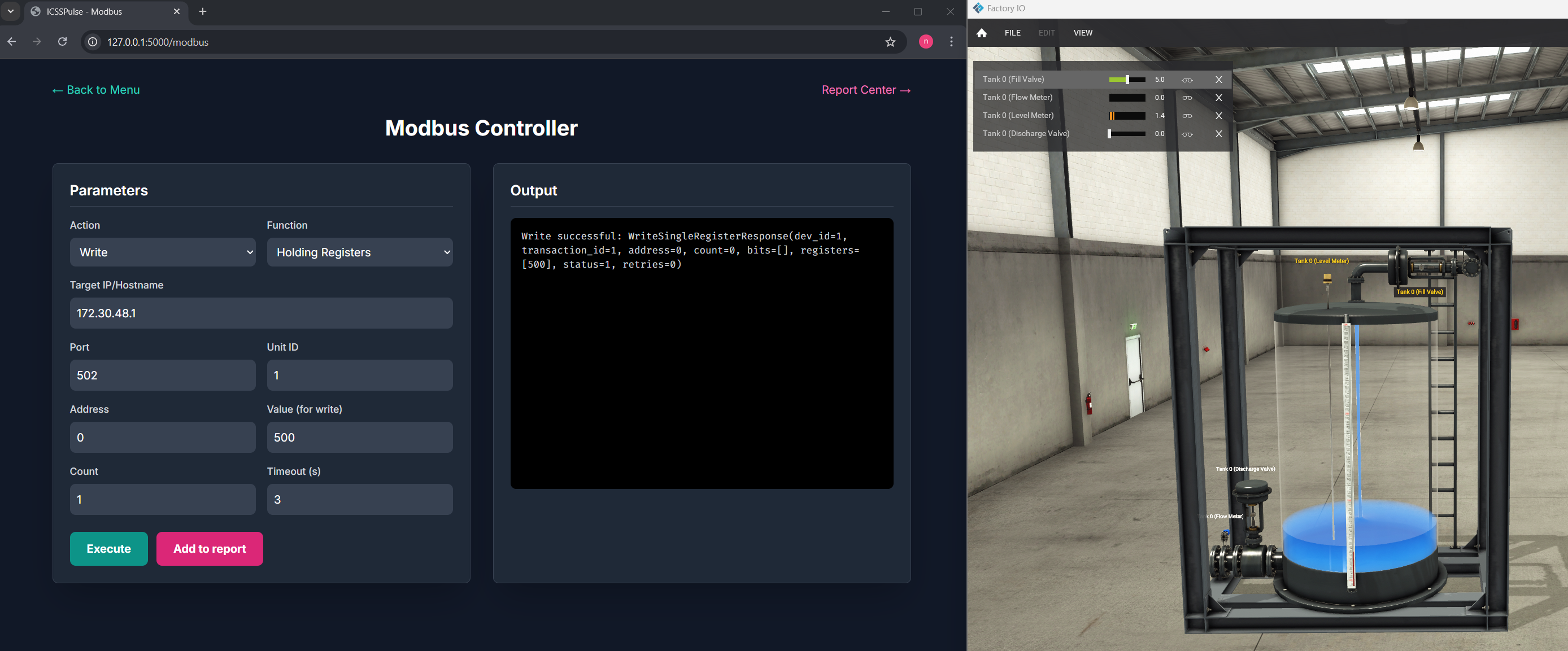}
\caption{Write operation using ICSSPulse.}
\label{fig:mw}
\end{figure}

Both testbeds were treated as black boxes, in accordance with the scope described in Section~\ref{S:Threat:Model}. The evaluation began with scan-based discovery, followed by manipulation of Modbus data types. In both scenarios, \textit{ICSSPulse} successfully identified active Modbus services, with the PyModbus server operating on port~5002 and the water treatment plant using the default Modbus port~502, confirming the presence of Modbus devices in the simulated OT network. Subsequently, the Modbus handler used to identify active unit IDs, enumerate all supported data types, including coils, discrete inputs, holding registers, and input registers, and determine the address ranges used to store the process values via a scan register range functionality. Finally, read and write operations were performed to manipulate register values. For example, writing an arbitrary value~500 to the fill valve register resulted in a value of~5 due to the water tank’s 100-scale configuration. An execution example is shown in Figure~\ref{fig:mw}.

Regarding the OPC~UA handler evaluation, we implemented a custom Python-based OPC~UA server using the \textit{python-opcua} library, as detailed in~\ref{SS:opcua}. Specifically, the server modeled an industrial production line organized through a hierarchical namespace structure. The exposed address space consisted of a root factory object (\textit{ns=2; i=1;}), representing the facility context, and a production line object (\textit{ns=2; i=2;}) corresponding to a specific unit. This production line contained three temperature sensors, i.e., node IDs 10–12, providing temperature measurements and two motors, i.e., node IDs 20–23, exposing motor speed and operational status (on/off). For our evaluation purposes, anonymous authentication was enabled, reflecting real-world deployments in which additional authentication mechanisms are not configured, which contradicts the principle of establishing secure defaults.

To evaluate the OPC~UA handler, we connected to the server using the endpoint \texttt{opc.tcp://localhost:4840/freeopcua/server/}. Initially, the handler correctly identified that the server accepted anonymous connections and operated under a \textit{None} security policy. After establishing the session, the \textit{Browse} functionality was used to traverse the hierarchical information model, revealing the complete namespace tree rooted at the \textit{Factory} object (\textit{ns=2; i=1;}) and its child nodes, including \textit{ProductionLine1}, \textit{TemperatureSensors}, and \textit{Motors}. Subsequently, the \textit{Enumerate} function was employed to profile available variables, distinguishing between read-only system metrics, such as \textit{Uptime} (\textit{ns=2; i=30;}), and writable process control variables. Importantly, the handler accurately reported node metadata, including data types (e.g., \textit{Double} for temperature sensors and \textit{Int32} for motor speed values). Finally, the evaluation concluded with active exploitation through read and write operations, retrieving real-time simulated sensor values, and demonstrating process control by modifying motor speed (\textit{ns=2; i=20;}) and motor status variables.

Last, after completing the Network Scan, Modbus, and OPC~UA assessment scenarios, we evaluated the ICSSPulse reporting capabilities, sending the results of each action to the LLM-powered reporting module. The module successfully consolidated the resulting network scan data, enumeration logs, and records of successful read/write exploitation into a unified context for feeding the GPT-4o-mini model. Consequently, the evaluation fulfilled the generation of two distinct report formats: a technical report mapping technical details of the findings, such as network ports, services, protocol data types, and corresponding mitigation strategies as derived from the ICS MITRE ATT\&CK matrix, and an executive summary translating these technical findings into high-level business risks regarding production continuity and safety.

\section{Conclusions}
\label{S:concl}

This work presents \textit{ICSSPulse}, an open-source, modular, and extensible PT platform for ICS protocols. To the best of our knowledge, \textit{ICSSPulse} is the first platform that unifies functions like network scanning, protocol-aware Modbus and OPC~UA interaction, and LLM-assisted reporting within a single, web-based, user-friendly PT ecosystem, using a lightweight Python-based backend combined with simulated industrial services, while preserving protocol fidelity and operational transparency. Through experimental evaluation on Modbus test servers, a \textit{Factory I/O} water treatment scenario, and a custom OPC~UA production-line model, we exhibit that \textit{ICSSPulse} can successfully discover industrial services, enumerate process-relevant assets, and perform unauthorized read and write operations that influence process behavior. A key novelty of the platform lies in its ability to translate low-level technical findings into structured executive and technical reports through an LLM-assisted module, exemplifying the practical integration of LLMs for automated reporting in ICS PT workflows. Despite these contributions, we acknowledge several limitations in terms of scalability and realism. Namely, \textit{ICSSPulse}'s reliance on Python-based execution, containerized auxiliary tools, and software-only, unauthenticated testbeds constrains performance and experimental fidelity, rendering the platform suitable only for small- to medium-scale laboratory environments. However, these limitations motivate several avenues for future work, including the integration of additional industrial communication protocols, the offloading of performance-critical components to lower-level implementations, and the extension of evaluations to authenticated deployments and hardware-in-the-loop testbeds.

\section*{Acknowledgments}

This work is supported by the Research Council of Norway through the SFI Norwegian Centre for Cybersecurity in Critical Sectors (NORCICS) project no. 310105.

%
%
%
\bibliographystyle{splncs04}
\bibliography{bibliography}
\end{document}